\documentstyle[prd,epsfig,aps]{revtex}
\newcommand{\ba}{\begin{eqnarray}}
\newcommand{\ea}{\end{eqnarray}}
\newcommand{\be}{\begin{eqnarray}}
\newcommand{\ee}{\end{eqnarray}}
\newcommand{\bi}{\bibitem}

\newcommand{\mplq}{m_{Pl}^2}

\newcommand{\bb}{\beta_1 \beta_2}
\begin{document}
\draft
\input epsf

\twocolumn[\hsize\textwidth\columnwidth\hsize\csname@twocolumnfalse\endcsname
\title{ Born-Infeld type Gravity  }
\author{\bf D. Comelli}
\address{     
{\it INFN - Sezione di Ferrara, 
via Paradiso 12, I-35131 Ferrara, Italy  }}
\date{May 2005}
\maketitle
\noindent

\vspace{0.3cm}
Generalizations of   gravitational   Born-Infeld type  lagrangians are investigated.
Phenomenological constraints (  reduction to Einstein-Hilbert  action for small curvature,
  spin two ghost freedom
and  absence of Coulomb like Schwarschild singularity )
 select one  effective lagrangian whose dynamics is dictated by the tensors
$g_{\mu\nu}$ and ${\cal R}_{\mu\nu\rho\sigma}$ (not $R_{\mu\nu}$ or the scalar $R$).
\vskip1pc
]

\vspace{2.cm}

\section{ Introduction.}~~

There are numerous suggestions in the literature for modification of the
classical Einstein Hilbert (EH) action of general relativity:    
\be
S_E = \frac{\mplq}{16\pi}\,\int d^4x\, \sqrt {-g}\, \left( R-2 \Lambda \right)
\label{se}
\ee
where $R$ is the curvature scalar, $g =  \det|| g_{\mu\nu}||$ is the determinant
 of the metric tensor, $\Lambda$ the cosmological constant and $m_{Pl}$ is the Planck mass. 
Many modification attempts (at least in 4-dimensional space-time) are based on the same structure of the
 action integral with the addition to the Einstein term of some scalar functions of $R$ and/or of combinations of
the Ricci, $R_{\mu\nu}$, and/or Riemann, ${\cal R}_{\mu\nu\alpha\beta}\,$ tensors
($ \int d^4x\, \sqrt {-g}\, {\cal L}(R,R_{\mu\nu}, {\cal R}_{\mu\nu\alpha\beta} )$).
Usually, but not necessarily, one considers quadratic in curvature terms 
proportional to $R^2$, $R_{\mu\nu}R^{\mu\nu}$, and ${\cal R}_{\mu\nu\alpha\beta}{\cal R}^{\mu\nu\alpha\beta}$.
Higher order or even non-local terms may appear as a result of quantum 
corrections, see e.g. the book\cite{bd}.
Such a form of the Lagrangian density, i.e. a scalar function multiplied by
the determinant of metric tensor, is dictated by the demand of  invariance of the action with 
respect to general coordinate transformation. 
However, this is not the only way to ensure this invariance.
In fact we can build many different unconventional invariants 
using the properties of the   Levi Civita tensor
\be
\epsilon_{\mu\nu\rho\sigma}=\sqrt{- g }\;e_{\mu\nu\rho\sigma}
\ee
where $e_{\mu\nu\rho\sigma}$ is the Levi Civita pseudo tensor (or permutation operator, $e_{0\,1\,2\,3}=1$ ),
 a tensorial density  of weight $w=1$ (note that $e^{\mu\nu\rho\sigma}$ is a tensorial density of
 weight $w=-1$).
\footnote{ The weight $w$ of a tensor density  $T^{\mu}_{\nu}$ is defined by the transformation law:
$T'^{\mu}_{\nu}=\left| \frac{\partial x}{\partial x'}\right|^w \frac{\partial x'^{\mu}}{\partial x^{\alpha} }
\frac{\partial x^{\beta}}{\partial x'^{\nu} }T^{\alpha}_{\beta} $ . }

The determinant of a second rank tensor can  be included as one of such  cases
\footnote{The usual determinant definition is:
$det||V_{\alpha \beta} ||=\frac{1}{4 !}e^{\mu_1\nu_1\rho_1\sigma_1}e^{\mu_2\nu_2\rho_2\sigma_2}
V_{\mu_1 \mu_2} V_{\nu_1 \nu_2}V_{\rho_1 \rho_2} V_{\sigma_1 \sigma_2}\,.  $
}.
It is easy to show that, taking  a generic two index tensor $V_{\mu\nu}$,
 we can build the following scalar densities of  weight  $w$:
\ba\nonumber
det||V_{\mu \nu} ||&&\rightarrow \;w=-2\\\nonumber
det||V^{\mu \nu} ||&&\rightarrow \;w=2\\\nonumber
det||V_{\mu }^{\nu} ||&&\rightarrow \;w=0\nonumber
\ea
In this spirit, Born and Infeld (BI) proposed for the electromagnetism the action \cite{bi,comment}
\be
 \int d^4x\sqrt{det||g_{\mu\nu}+\lambda\,F_{\mu\nu}||}
\ee
while,  Eddington  \cite{eddi} indicated,  as purely affine gravitational action, the term
\be
 \int d^4x\sqrt{det||R_{\mu\nu}(\Gamma)||}
\ee
whose possible generalizations are analyzed in ref.\cite{vollick}.

Taking into account only a purely  gravitational theory without any matter content
and considering a purely metric approach, we have at our disposal:
\begin{itemize}
\item one scalar $R=g^{\mu\rho}g^{\nu\sigma}  {\cal R}_{\mu\nu\rho\sigma}$
\item two (symmetric) tensors : $R_{\mu\nu},\;g_{\mu\nu}$
\item two  4-index tensors: $ {\cal R}_{\mu\nu\rho\sigma}$, $\epsilon_{ \mu\nu\rho\sigma}$ 
\end{itemize}
We can also  build some  4-index combinations of 
$R_{\mu\nu},\;$ and $ g_{\mu\nu}$,
with the same symmetry properties of $ {\cal R}_{\mu\nu\rho\sigma}$:
\ba
\hat {\tt G}_{\alpha_1\alpha_2 \bb}&=&
g_{\alpha_1\beta_1}\;g_{\alpha_2\beta_2}-g_{\alpha_1\beta_2}\;
g_{\alpha_2\beta_1}
\\
\hat {\rm R}_{\alpha_1\alpha_2 \bb}&=&
g_{\alpha_1\beta_1}\;R_{\alpha_2\beta_2}-g_{\alpha_1\beta_2}\;
R_{\alpha_2\beta_1}-\\&&\nonumber
g_{\alpha_2\beta_1}\;R_{\alpha_1\beta_2}+g_{\alpha_2\beta_2}\;
R_{\alpha_1\beta_1}
\\
\hat {\sf R}_{\alpha_1\alpha_2 \bb}
&=&
R_{\alpha_1\beta_1}\;R_{\alpha_2\beta_2}-R_{\alpha_1\beta_2}\;
R_{\alpha_2\beta_1}
\ea
All these 4-index tensors are not independent, in fact in four dimension:
\be
{\cal R}_{\mu\nu\rho\sigma}=\frac{1}{2}\; \hat {\rm R}_{\mu\nu\rho\sigma }-
\frac{R}{6}\;\hat {\tt G}_{\mu\nu\rho\sigma}+C_{\mu\nu\rho\sigma}
\ee
where $C_{\mu\nu\rho\sigma}$ is the Weyl tensor or Conformal tensor.

In ref.\cite{deser}, Deser and Gibbons proposed  the general covariant action :
\be S= \int d^4 x \sqrt{  det||g_{\mu\nu}+\lambda\,R_{\mu\nu}+X_{\mu\nu}|| }
\label{new}
\ee
where  $X_{\mu\nu}$ contains terms of second or higher order  in the curvature and 
formulated the minimum physical request that such a theory should satisfy:

1) Reduction to EH action for small curvature;

2) Ghost freedom;

3) Regularization of some singularities 
(for example the  Coulombian  like in the  Schwarzschild case);

4) Supersymmetrizability 
\footnote{   This requirement results quite stringent and 
is probably implemented if gravity descends from String/M-Theory \cite{woh,BIsusy,comment}. }.

All these points will be reanalyzed in chapter $V$ where a candidate lagrangian will be selected.

In ref.\cite{cd} there is an extensive analysis of the cosmological behaviors
 of actions like (\ref{new})  in  Friedmann Robertson Walker (FRW) background.

A straightforward generalization of  action (\ref{new}) can be written in the general form of  {\it ``determinant-action''}:
\be
S_{det}=\int d^4 x\sqrt{  det||  {\cal G}_{\mu\nu} (g_{\alpha\beta},\,R,\,R_{\alpha\beta},
\,{\cal R}_{\alpha\beta\delta\gamma} )|| }
\label{new1}
\ee
where ${\cal G}_{\mu\nu}$ is a  two index covariant tensor, combination of
$g_{\alpha\beta}$, $R$,  $R_{\alpha\beta}$ and  ${\cal R}_{\alpha\beta\delta\gamma}$.

Being $det||{\cal G} ||$ not necessarily positive definite,
$S_{det}$ can become imaginary in some portion of the $g_{\mu\nu}$ space.

A possible solution is given by matrices ${\cal G}$ that are  product of two
other matrices ${\cal M}$ and ${\cal N}$
\be
{\cal G}_{\mu\nu}=\!{\cal M}_{\mu}\,^{\alpha}
g_{\alpha\beta} {\cal N}_{\nu}\,^{\beta}
\ee
so that
\be det|| {\cal G}_{\mu\nu}||=g\;\; 
 det|| {\cal M}_{\mu}^{\nu}||\;det|| {\cal N}_{\mu}^{\nu}||
\ee
and in the case  ${\cal M}_{\mu\nu}={\cal N}_{\mu\nu}$
we have  $det||{\cal G}||=g\, det||{\cal M}||^2$ giving
\be
\int d^4x\; \sqrt{det||- {\cal G}_{\mu\nu}|| }=\int d^4x\;\sqrt{-g}\;\;| det||{\cal M}_{\mu}^{\nu}||\,|
\ee
and the action becomes automatically polynomial
\footnote{
In ref \cite{nieto}  the case ${\cal M}_{\mu}^{\nu}=\delta_{\mu}^{\nu}+
\lambda\,R_{\mu}^{\nu}$ with the lagrangian
$ \int d^4x   \sqrt{  det|| g_{\mu\nu}+2\lambda R_{\mu\nu}+\lambda^2 R_{\mu}^{\alpha}R_{\alpha\nu} ||} $
was studied.}.

Using the properties of the Levi Civita pseudo tensor $e$,
we can rewrite the  determinat-action (\ref{new1}) as:
\be
\int d^4 x\sqrt{  \frac{1}{4 \, !}\,\overline{ e \;e}\;
  {\cal G}_{\mu_1\nu_1}  {\cal G}_{\mu_2\nu_2} {\cal G}_{\mu_3\nu_3} {\cal G}_{\mu_4\nu_4}}
\label{new2}
\ee
where $\overline{ e \;e}$ is defined in appendix.
This form will be our guideline for the new generalizations of  BI type gravity in the next chapter.

\section{Generalized Born-Infeld Gravity}

The first generalization of the action (\ref{new2}) is obtained inserting 
an arbitrary number  ($n$)  of Levi Civita tensors:
\be
\int d^4x( \;\underbrace{ee}_2\; \;\underbrace{{\cal G}...{\cal G}}_4\;)^{1/2}\rightarrow
\int d^4x( \;\underbrace{e...e}_n\;\;\underbrace{{\cal G}...{\cal G}}_{2n}\;)^{1/n}
\ee
then, also the  ${\cal G}$ tensors, can be taken independents at each insertion: 
\be
\int d^4x\,( \underbrace{e...e}_n\;\;\underbrace{{\cal G}...{\cal G}}_{2n}\;)^{1/n}\rightarrow
\int d^4x\,( \underbrace{e...e}_n\;\;\underbrace{{\cal G}_1...{\cal G}_{2n}}_{2n}\;)^{1/n}
\ee

We stress that  $e^{\mu\nu\rho\sigma}$,
being a tensor density of weight $w=-1$ generates scalar densities that
 need to be corrected  taking the appropriate power $\frac{1}{n}$ for the full expression.
This allow us also to introduce directly into the lagrangian density 
higher index tensors, like $ {\cal R}_{\mu\nu\rho\sigma}$, generalizing the  determinant operation to
 tensors with more than two indices.

A generic term in $d$-dimensions, having $n$ times $e^{\mu\nu\rho\sigma}$,
$r$ times  $g_{\mu\nu}$, $s$ times  $R_{\mu\nu}$, and $t$ times
 ${\cal R}_{\mu\nu\rho\sigma}$ reads
\footnote{ We assume a sort of minimal  dimensional analysis  
not considering operators that saturate indices between them self,
 as  ${\cal R}_{\mu\nu\alpha\beta}{\cal R}^{\alpha\beta}\,_{\rho\sigma} $ and many others.
Each  single tensor  will saturate the respective indices only with
 the Levi Civita pseudo tensors (see fig.1 for $n=2$ and fig.2 for $n=4$).}
\be\label{count}
\int d^d x\;M^{d-\frac{2\,(t+s)}{n}}\;
(\;\underbrace{e...e}_n\;\underbrace{g...g}_r\;\underbrace{R...R}_s\;
\underbrace{\cal R...R}_t)^{1/n}
\ee
where the ``conservation of the number of indices'' requires 
$d\, n=2\,r+2\,s+4\,t$ and $M$ is a mass scale.
The range of variations of $t$ is : 
$0\leq t \leq d\, n/4$, (when $s+r=d\,n/2$ and $s=r=0$ respectively)
and the mass scale coefficient varies  from 
$M^{d-\frac{2\,s}{n}}$ to $M^{ \frac{d}{2} }$.

If we consider conformally flat spaces where $C_{\mu\nu\rho\sigma}=0$
we can simplify the general expression to
\be
\int d^d x\;M^{d-\frac{2\,s}{n}}\;
(\;\underbrace{e...e}_n\;\underbrace{g...g}_r\;\underbrace{R...R}_s\;)^{1/n}
\ee
where now $d\,n =2\,r+2\,s$. In this case $0\leq s \leq d\,  n/2$,
(for $ r=d\,n/2$ and  for $r=0$ respectively)
and the dimensionality of the mass coefficient runs from
$M^d$ for $s=0$  to $M^{0}$ for $s=2n$.

We note also that   actions with an odd number of $e\;$ 
violate  parity  and in the FRW or Schwarzschild backgrounds 
(that are our physical ``toy models''), these terms are exactly zero.

We will also reduce our analysis to the four dimensional space time, $d=4$,
leaving the extra dimensional spaces to future investigations.

Due to the fact the the number  of independent operators is growing very fast with
 $n$, we will concentrate on the  $n=2$ case and 
 discuss the generalizations with $n>2$ in general terms only. 

\begin{figure}[htb]
\epsfxsize=3.3 in
\epsfysize=0.6 in
\begin{center}
\leavevmode
 \epsfbox{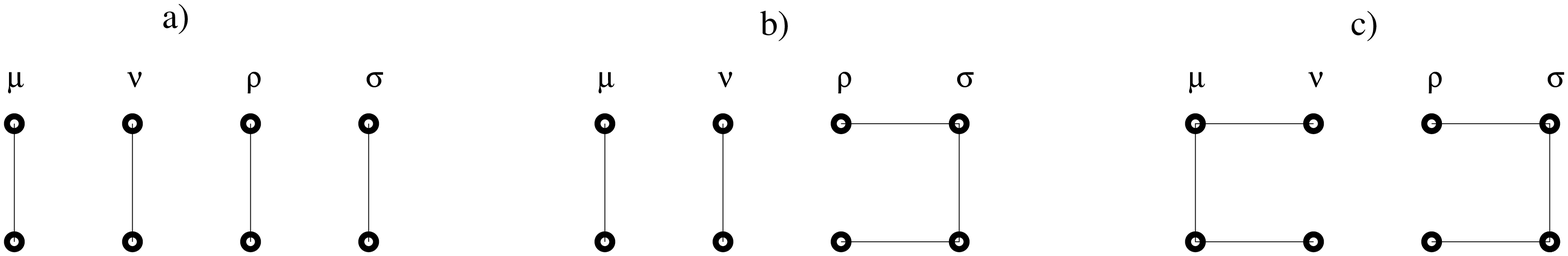}
\end{center}
\caption{Here we show the possible saturation patterns for the case $n=2$.
In $a)$ we saturate  with four 2-index symmetric tensors;
in $b)$ we saturate with two 2-index tensors and one 4-index tensor;
In $c)$ we saturate  with two 4-index tensors.
The  black points  indicate the four indices for each $e^{\mu_i\nu_i\rho_i\sigma_i}$ pseudo tensor
 (i=1,2).
}
\end{figure}

\subsection{Two $e$ }

Here we analyze of the case with only two Levi Civita pseudo tensors, $n=2$; 
the  counting rule of eq.(\ref{count}) fix the structure 
\be
\int d^4 xM^{4-t-s}\left. (\underbrace{ \overline{ee}}_{n=2}\underbrace{g...g}_{0\leq r\leq 4}
\underbrace{R...R}_{0\leq s\leq 4}
\underbrace{{\cal R}{\cal R}}_{0\leq t\leq 2} )^{1/2}
\right\arrowvert_{4 t+2r+2s=8} 
\ee
(in fig.1 we give a pictorial rule to saturate the various different indices )
from which, using the gravitational tensors at our disposal,
  we  generate the following operators of different dimensionality:

\subsubsection{\bf Operators of 0 dimension: $gggg$ (r=4)}
\be\label{n1}
 \overline{ e \;e}\; g_{\mu_1\nu_1}\!  g_{\mu_2\nu_2} \!g_{\mu_3\nu_3} g_{\mu_4\nu_4}  = 
\;*\hat {\tt G}\,*\hat {\tt G} \propto 
4 \,!\;g
\ee
where the $*$ product is defined in appendix.

\subsubsection{\bf Operators of dimension 2:  $gggR$ (r=3, s=1) and $gg{\cal R}$
(r=2, t=1) }
\ba
&&
\overline{ e \;e}\; g_{\mu_1\nu_1}\!  g_{\mu_2\nu_2} \!{\cal R}_{\mu_3\mu_4\nu_3\nu_4}  =2 
\;*\hat {\cal R}\,*\hat {\tt G}=4\, g\; R \\
&&
\overline{ e \;e}\;g_{\mu_1\nu_1}\!  g_{\mu_2\nu_2} \!g_{\mu_3\nu_3}
R_{\mu_4\nu_4} = \frac{1}{2}\, *\hat{\rm R}\,*\hat {\tt G}=6\, g\; R
\ea

\subsubsection{\bf Operators of dimension 4:  $ggRR$ (r=2, s=2), 
$\quad \quad gR{\cal R}$ (r=1, s=1, t=1) and ${\cal R}{\cal R}$ (t=2) }

\be \label{four1}
\overline{ e \;e}\, g_{\mu_1\nu_1}\!  R_{\mu_2\nu_2} \!
{\cal R}_{\mu_3\mu_4\nu_3\nu_4}  = 
* {\cal R}\, * \hat{\rm R}=
 \!2\,g\,( R^2-\!2[R]^2)
\ee

\be\label{four2}
\overline{ e \;e}\,{\cal R}_{\mu_1\mu_2\nu_1\nu_2} \!
{\cal R}_{\mu_3\mu_4\nu_3\nu_4}\!  =\!\frac{1}{4} \!\!
*\!{\cal R}\,*\!{\cal R} 
\!=\!\frac{g}{4} \!(\!{\cal R}^2\!-\!4[R]^2\!+\!R^2\!)
\ee

\be\label{four3}
\overline{ e \;e}
\,  g_{\mu_1\nu_1}  g_{\mu_2\nu_2} R_{\mu_3\nu_3}R_{\mu_4\nu_4}\!=
\!\frac{1}{4} \! *\!\hat{\rm R}\,*\!\hat {\rm R}\!=\!2 g\,(\!R^2\!-\![R]^2)
\ee
where the combination
$ ({\cal R}^2-4[R]^2+R^2) \equiv {\cal E}$ is the Gauss-Bonnet term.

\subsubsection{\bf Operators of dimension 6: $gRRR$  (r=1, s=3)  and
 $RR{\cal R}$ (s=2, t=1)  }

\ba\nonumber
&&
\overline{ e \;e}
\;  R_{\mu_1\nu_1}  R_{\mu_2\nu_2} {\cal R}_{\mu_3\mu_4 \nu_3\nu_4}=
 *{\cal R}\,*\hat {\sf R}=4\,g\,(
[ R \,{\cal R}\, R]+\\&&
 \quad\quad\quad\quad\quad\quad\frac{1}{2}R^3 -\frac{5}{2}R\, [R]^2+2[R]^3)
\\&&\nonumber
\overline{ e \;e}\;  g_{\mu_1\nu_1}  R_{\mu_2\nu_2} 
R_{\mu_3\nu_3} R_{\mu_4\nu_4}=\frac{1}{4}\; *\hat {\rm R}\,*\hat {\sf R}
=4\, g\;(R^3-
\\ &&\quad\quad\quad\quad \quad\quad 3 R\,[R]^2+2 [R]^3)
\ea

\subsubsection{\bf Operators of dimension 8:  $RRRR$ (s=4)}

\be\label{nl}
\overline{ e \;e}\;  R_{\mu_1\nu_1}  R_{\mu_2\nu_2} 
R_{\mu_3\nu_3} R_{\mu_4\nu_4}
= \frac{1}{4}\;*\hat {\sf R}\,*\hat {\sf R}
\ee

\begin{figure}[htb]
\epsfxsize=3.2 in
\epsfysize=1.4 in
\begin{center}
\leavevmode
 \epsfbox{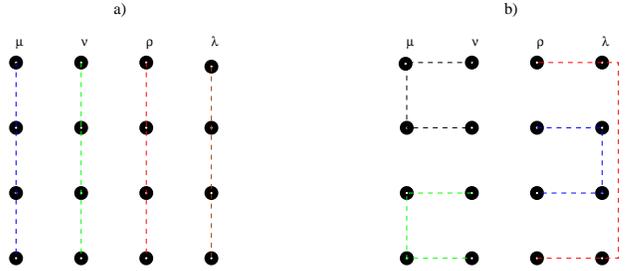}
\end{center}
\caption{The  black points  represent the indices of the four 
$e^{\mu_i\nu_i\rho_i\lambda_i}$ pseudo tensors
 (i=1,2,3,4) while the colored dashed curves are the way 
each ${\cal G}_{\mu_i\nu_i\rho_i\lambda_i}$ tensor
 saturates the respective $e$-tensors.
The first case $a)$ is describing eq.(\ref{det1}) while the case $b)$ 
is for eq.(\ref{det2}).
}
\end{figure}

\subsection{ Four $e$}

When there are four $e^{\mu\nu\rho\sigma}$ pseudo tensors, 
the large number of  possible saturations generates many invariants.
 The study or  the classification of all of them is beyond the scope of the present letter.
My purpose is  a  short analysis  of two possible choices that, from my point of view, are the
{\it most symmetric} ones. 

We define respectively a  `` determinant '' action for a 4-index tensor as
\ba\label{det1}\nonumber
\int d^4x \left(\,{\rm det} ||  {\cal G}_{\mu\nu\rho\sigma}||\,\right)^{1/4}
\equiv \int d^4x  \left(\overline{e\,e\,e\,e}\;{\cal G}_{\mu_1\nu_1\rho_1\sigma_1} \right.  
\\ \left.
{\cal G}_{\mu_2\nu_2\rho_2\sigma_2}\,
 {\cal G}_{\mu_3\nu_3\rho_3\sigma_3}\,
 {\cal G}_{\mu_4\nu_4\rho_4\sigma_4}\,\right)^{1/4}
\ea
where $\overline{e\,e\,e\,e}$  is defined in appendix, and
another interesting saturation pattern   given by
\be\label{det2}
\int d^4 x({\it det}|| {\cal G}_{\alpha \beta \gamma \sigma}||)^{\frac{1}{4}}\!\equiv \!
 \int \!d^4 x(Tr[*{\cal G}\!*{\cal G}\!*{\cal G}\!*{\cal G}])^{\frac{1}{4}}
\ee
 already investigated  in the paper \cite{woh}
with ${\cal G}=(1+k\,R)\,\hat{\tt G}$ $+\hat{\rm R}+\hat{\sf R}+ $ higher derivative terms.

In fig.2 it is shown the  saturation path corresponding respectively to
eq.(\ref{det1}) and to eq.(\ref{det2}).

In order to understand the different physical implications  from these two choices,
in section III $\, B$, we will analyze the respective actions obtained working with 
 ${\cal G}=M^2\hat{\tt G}+{\cal R} $ in a  FRW metric background.
 This simple exercise can give the feeling  of physical implications coming from
  ``higher order'' BI generalizations.

\section{Small curvature lagrangian expansion}

In this chapter we will discuss the small curvature  expansion of the most general 
 gravitational lagrangian obtained combining the  eqs. (\ref{n1}-\ref{nl})  
for $n=2$ while for  $n=4$ we discuss  only the special case with
  ${\cal G}=M^2\,\hat{\tt G}+{\cal R}$ in FRW space.

\subsection{Two $e$ case}

The most general combination of eqs. (\ref{n1}-\ref{nl}) gives
\ba
&&M^8(1+ \frac{\alpha_1}{M^2}\;R+
\frac{1}{M^4}(\beta_1\,R^2+\beta_2\,[R]^2+\beta_3\,{\cal R}^2)
+\\&&\nonumber
\quad\quad\quad \frac{1}{M^6}(\lambda_1\,R^3+ \lambda_2\,R\,[R]^2+...)+
\frac{\gamma}{M^8}(det||R_{\alpha\beta}||)
\ea
Where $ M $  is a dimensional mass coefficient
and   $\alpha_1,\,\beta_i,\,\lambda_i$ and $\gamma$ are  free  parameters.

In order to have some insight about the asymptotic small curvature
behavior we can expand around the leading operator, meaning the 
operator with the lowest dimensionality.

If for example the leading operator is the dimension zero one, 
then the  expanded lagrangian becomes 
\ba
\int d^4x \sqrt{-g} \left( M^4+\frac{\alpha_1}{2}\,M^2 \,R+
\left((\frac{\beta_1}{2}-\frac{\alpha_1^2}{4}
)R^2+ \right.\right.\\\left.\left.
\frac{\beta_2}{2}\,[R]^2+
\frac{\beta_3}{2}\,{\cal R}^2
  \right)+...\right)\nonumber
\ea
and if  we want to take care of the ghost problem
we need the constraint $-4\beta_3= \beta_2$ (see section $V$).

In the case the  leading operator is the two dimensional one,  we get
\ba
\!\!\int\!\! d^4\!x\! \sqrt{-g}   \left(\!\!\! M^3 \!\sqrt R+\!
\frac{M}{2 \,\sqrt R}\!\left(\beta_1R^2\!+\!
\beta_2 [R]^2\!+\!\beta_3{\cal R}^2\!
\right)\!+\!\!...\! \right)\!\!\!
\ea
For leading dimension four operator,
 in order to have the leading EH term, we need $\beta_2\!=\!\beta_3\!=\!0$ so that
\footnote{This cancellation is obtained properly combining the two operators of eq. (\ref{four1},\ref{four3}):
$\quad *({\cal R}-\frac{1}{2}\, \hat{\tt R})*\hat{\tt R}=*(-\frac{R}{6}\, \hat{\tt G}+C)*\hat{\tt R} $}
\ba
\int d^4x \sqrt {-g}  && \left( M^2 |R|+\frac{1}{2}\left(
\lambda_1R^2+\lambda_2[R]^2+\right.\right.
\\&& \nonumber\left.\left.\lambda_3\frac{[R]^3}{R}+
\lambda_4\frac{[R {\cal R}R ]^3}{R}\right)+...  \right)
\ea
etc...

In order to get a phenomenological acceptable model
we will follow the  usual attitude  to cancel ``by hand'' the
cosmological constant with the following trick
\be
\int d^4x\sqrt{-g}M^4\;\left( \sqrt{(1+\frac{\alpha}{M^2}\;R
+\frac{1}{M^4}\;...)}-1\right)
\ee
with no justification from the symmetry point of view.

 \subsection{Four $e$ case}
 
As already discussed before we decided to 
work out this case with two  models in FRW metric.
Taking  ${\cal G}=M^2\,\hat{\tt G}+{\cal R}$  the  small curvature expansion is
 around a non zero cosmological constant value for both the eqs. (\ref{det1},\ref{det2}).
The next to leading operators are for eq.(\ref{det1}) 
\be
\int d^4x \sqrt{-g}\,M^2\overline{e\,e\,e\,e}\,
\hat{\tt G}\,\hat{\tt G}\,\hat{\tt G}\,{\cal R} \sim\!\!\!\int dt\,a^3 M^2 \left(3\,H^2+H' \right)=0
\ee
while for eq.  (\ref{det2})
\be
\int d^4x \sqrt{-g}\, M^2\,
\hat{\tt G}*\hat{\tt G}*\hat{\tt G}*{\cal R}\sim \int d^4x \sqrt{-g}\,M^2\, R
\ee
The next to next to leading  operators correspond to
\be
\int d^4x \sqrt{-g}\; \hat{\tt G}\,\hat{\tt G}\,{\cal R }\,{\cal R} \sim \int dt\,a^3\; H^4
 \ee
from eq.(\ref{det1}) while eq.(\ref{det2}) generates
\be
\int d^4x \sqrt{-g}\;\;Tr[ *\hat{\tt G}*\hat{\tt G}*{\cal R}*{\cal R}]=\int d^4x \sqrt{-g}\, {\cal E}=0
\ee

It is evident that for this particular choice of ${\cal G}$
 eq.(\ref{det1}) fails to match EH at small curvature and predict ghosts (see next chapter),
while eq.(\ref{det2}) fit the EH constraint and results ghost free (see next chapter),
 being in such a way  an interesting phenomenological theory of gravity (see \cite{woh}).

\section{ Symmetries}

The existence  of some guiding symmetry principle to build  our action
  will strongly improve the prediction of our approach. 
We know for example  that the effective gravity lagrangians derived by String Theory 
to the fourth  order in curvature expansion
are ghost free \cite{stringhost}.
This it means that in second curvature order we have the Gauss Bonnet combination (${\cal E}=R^2-4[R]^2+{\cal R}^2$).
In higher order we find cubic corrections for bosonic string theories while the supersymmetric extensions 
predict zero cubic and non zero quartic corrections  \cite{string}.
Because string inspired effective lagrangians are computed evaluating on shell graviton amplitudes
 ($R_{\mu\nu}=0$), the terms which contains at least twice the Ricci tensor or the Ricci scalar are non properly included.

Supersymmetric BI type generalizations of Weyl supergravity action are given in \cite{BIsusy}.

Since there is no decisive hint about the correct string model, 
we will attempt a phenomenological approach guessing
 some ad hoc principles and testing the possible implications.

Here we will introduce  a sort of ``{\it selection rule}''
such that  only some   tensors generate the gravitational dynamics.
Our possible choices  are in between $\hat{\tt G},\;\hat{ R},\;\hat{\sf R}$ and ${\cal R}$.

If for example, only the tensor ${\cal R}_{\mu\nu\rho\sigma} \;$ ($t=2$)
is present, the lagrangian  results:
\be
\int d^4x \sqrt{ -g}\;M^2\;\sqrt{{\cal R}^2-4[R]^2+R^2}
\ee

In the case with only the tensors
$\hat {\tt G}_{\mu\nu\rho\sigma}$ and ${\cal R}_{\mu\nu\rho\sigma} \;$ 
 $((r=4)+(r=2, t=1) + (t=2))$ we get:
\be\int d^4x \sqrt{-g}&M^4&\,
\sqrt{ *\hat {\tt G}*\hat {\tt G}+\frac{\alpha'}{M^2} 
*\hat {\tt G}*\!{\cal R}+\nonumber
\frac{\beta'}{M^4}\,*\!{\cal R}*\!{\cal R}}=\\
\int d^4x \sqrt{-g}\;&M^4&\,\sqrt{ 1+\,\frac{\alpha''}{M^2} \,R+
\,\frac{\beta''}{M^4}\,{\cal E}}
\ee
and it fits exactly the constraints for ghost freedom and the EH asymptotic (see cap.$V$).

While in the opposite case where no ${\cal R}_{\mu\nu\rho\sigma} $ can enter
 $((r=4)+(r=3, s=1) + (r=2, s=2)+(r=1, s=3) + ( s=4))$,
we have  the lagrangian:
\be\label{mio}
\int d^4x \sqrt{-g}\;M^4\,\sqrt{ 1+ \frac{\alpha}{M^2}\,R+
\frac{\beta}{M^4}\,(R^2-[R]^2)+}\\
\overline{\frac{\delta}{M^6}\,(R^3-3\,R\,[R]^2+2\,[R]^3)+\frac{\gamma}{M^8}\,det||R_{\mu\nu}||}\nonumber
\ee
where ghosts  show up (see cap.$V$).

When only ${\cal R}_{\mu\nu\rho\sigma} $ and
${ R}_{\mu\nu} $ tensors are present $(( t=2)
+ (t=1, s=2)+ ( s=4))$ :
\be
\int d^4x \sqrt{-g}&&\;\sqrt{ M^4\,({\cal E})+
M^2\,(...)+\gamma\,det||R_{\mu\nu}||}
\ee

Finally we have   only one  operator with  ${ R}_{\mu\nu} $ tensors 
$\;( s=4)$ that turns out to be also local Weyl invariant:
\be
\int d^4x \sqrt{det||R_{\mu\nu}||}
\ee
with no EH matching.

\section{ Phenomenological Model}

The only phenomenologically interesting lagrangian coming from the above selection rules is
certainly
\footnote{Note that the  expression under square root corresponds to the  Lovelock lagrangian in four dimensions \cite{lov}. }
\be\label{my}
\int d^4x\sqrt{-g}M^4\!\left(\!1\!-\! \sqrt{1\!-\!\frac{\alpha\,R}{M^2}
+\frac{\beta}{M^4}({\cal R}^2-4[R]^2+R^2)}\;\right)
\ee
where we have applied the cancellation mechanism of cosmological constant.

At this point, we will reanalyze more carefully the above lagrangian 
to the light of the physical criteria suggested by Deser and Gibbons 
in \cite{deser}.

\subsubsection{ EH small curvature limit}

The small curvature limit of eq.(\ref{my}) results:
\ba\int &d^4x& \sqrt{-g}\,\left(\nonumber
\alpha \frac{M^2}{2}\,R+\frac{\alpha^2}{8} \,R^2-
\frac{\beta}{2} ({\cal R}^2-4\,[R]^2+R^2)+\right.
\\\nonumber
&\frac{\alpha\,R}{16  M^2}&  \left.\left((\alpha^2-4 \beta)R^2-4\,\beta({\cal R}^2-4\,[R]^2 ) \right)+...\right)=
\\
\int &d^4x& \sqrt{-g}\,\frac{\mplq}{16 \pi}\left(\,R+ \frac{1}{6\,m_0^2}R^2+...\right)
\ea
So, the requirement of a correct EH leading  gravitational operator fix $\alpha \frac{M^2}{2}= \frac{m_{pl}^2}{16 \pi}$
while  the coefficient of the $R^2$ operator results
$\frac{\alpha^2}{8}= \frac{m_{pl}^4}{512 \pi^2\,M^4}$ 
generating  an extra scalar degree of freedom  with mass 
$m_0=\sqrt{\frac{16 \pi}{3}} \frac{M^2}{m_{Pl}} $.
The exchange of such a scalar between two test particles changes the $1/r$ static gravitation potential slope 
to $\frac{1}{r}\left(1 +\frac{1}{3} e^{-m_0\,r}\right)$.
Using the results of ref.\cite{pot} with the strength parameter equal to $1/3$ 
  we can obtain a lower bound on $m_0$ of $\sim 2 \;10^{-2} eV$ corresponding to a value for 
the mass parameter $M \geq$  250 GeV.

\subsubsection{ Ghost Cancellation}

The  analysis of the particle content
in higher derivative lagrangians of the form 
\be
\int d^4x\;\sqrt{-g}\;\; F\!\left[R,[R]^2,{\cal R}^2\right]
\ee
shows the existence of massless gravitons plus new degrees of freedom.
 In general there is   a massive spin zero field  ($m_0$ mass) and  a massive spin two field ($m_2$ mass)
 with a wrong sign of kinetic term: a ghost fields.
Due to  the fact that the mass of such  particles and
 their potential ghostlike may be very different around different vacuum states
we give the full set of eqs that fix such a parameters
around solutions 
characterized by a  constant curvature $R=R_0$ in a maximally symmetric background \cite{spectrum},
${\cal R}_{\mu\nu\rho\sigma}=\frac{R_0}{12}(g_{\mu\rho}\,g_{\nu\sigma}-g_{\mu\sigma}\,g_{\nu\rho})$.

The equations of motion that fix $R_0$ are:
\be\label{eqm}
\left.
F-\frac{R}{2}F_{R}-\frac{R^2}{4}F_{P}-\frac{R^2}{6}F_{Q}\right|_{R_0}=0
\ee
where  $P\equiv [R]^2$ and $Q\equiv {\cal R}^2$ and  $F_R=\frac{\partial F }{\partial R},\;
F_P=\frac{\partial F }{\partial P},\;F_Q=\frac{\partial F }{\partial Q},\;F_{PP}=\frac{\partial^2 F }{\partial P^2}
$ and so on.

To each solution of  (\ref{eqm}) it corresponds a cosmological constant \cite{spectrum} given by
\be
\Lambda=-\frac{8 \pi}{m_{Pl}^2}\left.\left( F-F_R\;R+\frac{1}{2} F_{RR}\;R^2\right) \right|_{R_0}
\ee
with an effective Planck mass
\be
\frac{m_{Pl}^2}{16\pi}=\left. \left(F_R-F_{RR}\;R\right) \right|_{R_0}
\ee
and the masses of the two extra degrees of freedom $m_0$ and $m_2$: 
\ba
\frac{m_{Pl}^2}{96 \pi\,m_0^2}&=&\frac{1}{2}F_{RR}\nonumber
+\frac{1}{3}(F_{P}+F_{Q})+\frac{R}{6}(3F_{RP}+2F_{RQ})+\\
&&\left.R^2(\frac{1}{8}F_{PP}+\frac{1}{6}F_{PQ}+\frac{1}{18}F_{QQ})
\right|_{R_0}
\ea
\be
-\frac{m_{Pl}^2}{32 \pi\,m_2^2}=
 \left.\frac{1}{2}F_{P}+2F_{Q}\right|_{R_0}
\ee
 where we always take $R=R_0$ and  $P=\frac{R_0^2}{6},\;$ 
$Q=\frac{R_0^2}{4}$.

A ghost free spectrum (as required for example from string theory 
\cite{stringhost}) is realized when $m_2 \rightarrow \infty$ and
 this request is automatically satisfied  for lagrangians of the form
\be
F\!\left[R,{\cal R}^2-4 [R]^2\right]= f\!\left[R,{\cal E}\right]=
f\!\left[(*\hat{\tt G}*{\cal R}),\,(*{\cal R}*{\cal R})\right]
\ee
that fit the structure of eq.(\ref{my}).

For this particular lagrangian (\ref{my})  we have two possible background solutions: 
one flat background $R_0=0$ with zero cosmological constant,
 that corresponds to the small curvature EH limit previously described;
one with $R_0\neq 0$  (we can give analytical results only in the small $\alpha$ limit
  corresponding to $m\geq m_{Pl}$ :
$m_{Pl}^2=8\pi\alpha\,m^2+O(\alpha^3),\;$
$R_0\sim \frac{3}{40 \pi \beta}m_{Pl}^2+O(\alpha^3),$ 
 a non zero cosmological constant $\Lambda\sim \frac{4.8 \: 10^{-7}}{m^4\,\beta^2}m_{Pl}^6+O(\alpha^3)$,
 and an extra scalar degree of mass  $m_0^2\sim \frac{25\pi m^4}{3 m_{Pl}^2} +O(\alpha)$).
In fig.3 we plot in $(\alpha,\,\beta)$ parameter space   the zero limit for $m_0^2$ and $\Lambda$.
 Only the region  $m_0^2>0$ is stable, while $\Lambda$ can have both signs
 \cite{aros}(note that only positive $\alpha,\,\beta$ values are allowed).

\begin{figure}[htb]
\epsfxsize=3.1 in
\epsfysize=2.5 in
\begin{center}
\leavevmode
 \epsfbox{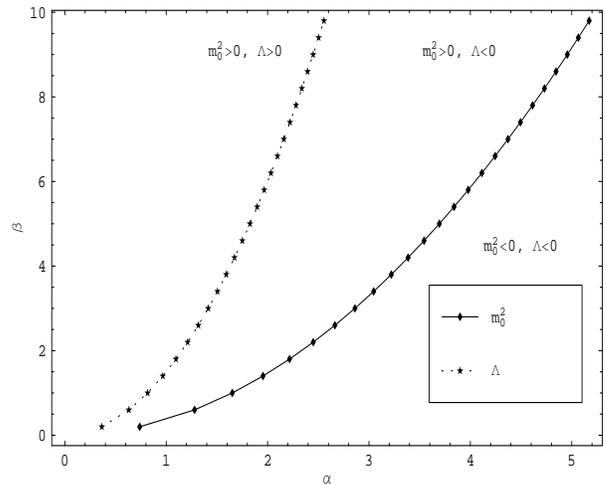}
\end{center}
\caption{
$\alpha-\beta$ parameter space for  $m_0$ (straight line) and $\Lambda$ (dot line) zero values, 
in the case with $R_0\neq 0$.}
\label{absmatt}
\end{figure}
\vspace{0.1cm}

\subsubsection{ Schwarzschild singularity}

\vspace{0.1cm}

Many generalizations of EH action induce corrections to  Schwarzschild metric
 which could have interesting consequences.
In the class of Born-Infeld action there are  some  studies about spherically symmetric
 Schwarzschild solutions (see \cite{woh,bh}) .
In this paper we are not interested in a full analysis 
of   new solutions because we can reduce our action to the one studied in
\cite{bh} in some portion of parameter space.

Following the lines of \cite{bh}, we can  neglect, in the action,
 terms proportional to $R$ and $R_{\mu\nu}$, due to that fact that we are looking for  solutions  similar to the
 Schwarzschild one ( $R_{\mu\nu}\sim 0$).
 Only the presence  of terms proportional to the Weyl tensor can, in principle,
 remove the black hole singularity at the origin.
This observation reduce our action to
\be
\int d^4x\sqrt{-g}\,M^4\,\left( 1-
\sqrt{1+\frac{\beta}{M^4}\;{\cal R}^2}  \;\right)
\ee
and this exact form is studied in chapter four of \cite{bh} (see there for details)
\footnote{Note that in this case we need a negative  $\beta$ parameter 
that cancel the stability of the background $R_0\neq 0$ solution (see chapter V 2) 
leaving only the $R_0=0$  background. }.
The main results in \cite{bh} are the existence of solutions which behave  asymptotically
as black holes and becomes spaces of constant ${\cal R}^2$ at small radii.
In some portion of  parameters space there is not even an  event horizon with
 the presence of a bare mass instead of a black hole
(bare in the sense that is not hidden behind an event horizon) 
without a naked singularity.

\vspace{0.1cm}

\section{ Conclusions}

\vspace{0.1cm}

In this paper we have discussed some generalization of determinant gravity
following the  steps:
\ba \nonumber
\int d^4x\;\sqrt{ det||{\cal G}_{\alpha\beta} ||}&=&
\int d^4x( \;\underbrace{ee}_2\;
\;\underbrace{{\cal G}...{\cal G}}_4\;
)^{1/2}\rightarrow\\\nonumber
&&\int d^4x( \;\underbrace{e...e}_n\;
\;\underbrace{{\cal G}...{\cal G}}_{2n}\;
)^{1/m}\rightarrow
\\\nonumber
&&\int d^4x( \;\underbrace{e...e}_n\;
\;\underbrace{{\cal G}_1...{\cal G}_{2n}}_{2n}\;
)^{1/n}\rightarrow
\\\nonumber&&
\int d^4 x 
(\;\underbrace{e...e}_n\;\underbrace{g...g}_r\;\underbrace{R...R}_s\;
\underbrace{\cal R...R}_t)^{1/n}
%
\ea
with $4 n= 2 r +2 s+4 t$.

Then we analyzed all the possible operators  obtained in the case $n=2$.
Selecting as guide lines the following requests:

1) Reduction to EH action for small curvature;

2) Ghost freedom;

3) Regularization of some singularities;

4) Supersymmetrizability.

We selected the lagrangian
 \be\label{ph}\nonumber 
\int d^4x\sqrt{-g}\,M^4\,\left(1\!-\! \sqrt{1-\frac{\alpha\,R}{M^2}
+\frac{\beta}{M^4}({\cal R}^2-4[R]^2+R^2)}\;\right)
\ee
which has a EH leading term in the small curvature limit, 
it results ghost free and 
for some parameter space show indications for the cancellation of the Coulomb like Schwarzschild singularity.

For what concern the possible terms with $n\geq 4$, 
there are many interesting ``determinants'' definitions that have to be
physically investigated and to cover all of them requires some more effort.

\vspace{1.cm}
{\bf Acknowledgements}

I would like to thanks A. Dolgov,  A. Riotto and in particular M. Pietroni  for stimulating discussions.

\section{\bf appendix}
Some definitions:
\ba \label{def1}
&&
e^{\mu_1\mu_2\mu_3\mu_4}\,
e^{\nu_1\nu_2\nu_3\nu_4}\equiv\overline{ e \;e}\\
&&
e^{\mu_1\mu_2\mu_3\mu_4}\,
e^{\nu_1\nu_2\nu_3\nu_4}\;
e^{\rho_1\rho_2\rho_3\rho_4}\,
e^{\sigma_1\sigma_2\sigma_3\sigma_4}\equiv\overline{ e \;e\;e\;e}\\
&&
[R]^2\equiv R^{\mu\nu}R_{\mu\nu}\,,\;\;\;\;\;\;
[ R \,{\cal R}\, R] \equiv R^{\mu\rho}{\cal R}_{\mu\nu\rho\sigma} R^{\nu\sigma}
\ea

* product definition (here  applied to ${\cal R}$ tensor, analogous 
expressions result for the other 4-index tensors):
\be\label{def2}
\frac{1}{2}\,\epsilon^{\mu\nu\alpha\beta}\;
{\cal R}_{\alpha\beta\rho\sigma }\equiv
*{\cal R}^{\mu\nu}\,_{\rho\sigma }
\ee
\be
\frac{1}{2}\,{\cal R}_{\rho\sigma \alpha\beta }\;
\epsilon^{\alpha\beta\mu\nu}\equiv
{\cal R}\!*_{\rho\sigma }\,^{\mu\nu}
\ee

\be\label{def3}
\frac{1}{4}\,\epsilon^{\mu\nu\alpha\beta}\,
{\cal R}_{\alpha\beta\delta\gamma }\;
\epsilon^{\delta\gamma\rho\sigma}\equiv
*{\cal R}\!*^{\mu\nu \rho\sigma}
\ee


\end{document}